# Contiguous Hetero-structures & Co-existing Morphological Derivatives of Preferentially Grown Carbo-nitrides in Long Term Aged SS 316LN with Varying Nitrogen Concentration

Alphy George[*], Vaishnavi Krupa B. R., R. Mythili, Arup Dasgupta, J. Ganesh Kumar and G.V. Prasad Reddy

*Metallurgy and Materials Group, Indira Gandhi Centre for Atomic Research, HBNI, Kalpakkam-603102, India*

E-mail: alphy@igcar.gov. In

**Abstract:** To assess long-term structural integrity under operational conditions in nuclear reactors, SS 316LN with varying nitrogen content is subjected to thermal aging for 20000 hours at 650°C. While the annealing twins and heterogeneous grain evolution by bimodal division are independent of chemical composition, the type, size and morphology of evolving secondary phases are characteristic for nitrogen concentration. Colour contrast in bright field optical microscopy combined with transmission electron microscopy reveals the presence of three major precipitate phases; $M_{23}(C, N)_6$, $Fe_2Mo$ intermetallics and $Cr_2N$. Contiguously formed hetero-structures of $Fe_2Mo$ and $Cr_2N$ are quantified and extensive formation has been observed in steels with high nitrogen content (0.14 wt. % and 0.22 wt. %). Nickel enrichment in intermetallic, and cation/anion replacement by manganese/carbon in $Cr_2N$ are observed. In addition to the grain boundary carbo-nitrides, a unique occurrence of morphological derivatives of intra-granular $M_{23}(C, N)_6$ precipitates with cube-on-cube orientation relationship such as (1) isolated cubes, (2) primary stringers/strings of cubes, (3) secondary stringers/clusters of branched primary strings, and (4) clusters of laths are identified. The stringers and laths are preferentially grown in $\{111\}/\langle110\rangle\gamma$ and have crystallographic variants. All varieties of $M_{23}(C, N)_6$ are prominent in the sample with 0.22 wt. % nitrogen whereas, clustering of stringers and laths are absent in steels with lower nitrogen concentration. The conclusions are drawn from bright field colour contrast optical micrographs in the present study demonstrate the potential of light optical microscopy to be used as a significant tool to efficiently and accurately process huge amount of data including the preferential growth planes/directions ($\{111\}/\langle110\rangle$) of morphological derivatives of $M_{23}(C, N)_6$ precipitates, by analyzing the symmetry and dense planes/directions of secondary phases. Overall mechanical properties have been reformed by the precipitation strengthening effect arising from the excess nitrogen content.

## 1. Introduction

The development of high nitrogen steels targets improvement of mechanical properties such as strength, toughness, creep resistance and stress corrosion resistance of engineering components thus ensuring their extended and economic utility [1]. With greater solid solubility than carbon, nitrogen-a strong austenite stabilizer acts as an interstitial solid solution strengthener. It is reported that yield strength of nitrogen added steel exceeds standard carbon-based steel which can be further improvised by cold working [2, 3]. Currently, in fast breeder nuclear reactors, SS 316LN is used as the structural material for primary side components [4]. Nitrogen addition can enhance the resistance to high temperature fatigue and creep failure of the components experiencing cyclic thermal stresses, steady-state loading and thermal transients in addition to the mechanical load variations during normal reactor operations [5-8].

It has been well established that for SS 316LN, microstructures like grain morphology and twin density are least affected by heat treatment/prolonged annealing [9]. However, long term thermal aging promotes the formation of thermally activated secondary phases containing both interstitial strengtheners causing a reduction in high temperature stability of the alloy [2, 10]. While long-term aging of SS 316LN, the precipitate phases such as carbides, σ, η and $Cr_2N$ are expected to form in the matrix [11-14]. The chemistry and distribution of $M_{23}C_6$ carbides in SS 316LN aged for 300 hours has been well documented previously and chromium depleted regions are quantitatively measured [15]. On the other hand, though $Cr_2N$ precipitation processes have been studied extensively, there are major discrepancies remain in explaining morphology and distributions of the secondary phases in the matrix.

Structural components using 316LN SS expose to high temperature for prolonged periods of 40-60 years in typical sodium cooled fast breeder reactor that operates at 550ºC. Hence it is important to study the evolution of microstructural phases in thermally aged specimens in detail for a greater understanding of the mechanical behavior of the material. As a part of these investigations on long-term microstructural stability, the authors target the morphological and structural changes of secondary phases in long term aged SS 316LN with varying nitrogen concentration.

## 2. Materials and Methods

Chemical composition of four heats of 316LN SS with 0.07, 0.11, 0.14, and 0.22 wt %N (hereafter designated as N07, N11, N14 and N22 respectively) is given in Table 1. Rectangular blanks of dimensions, 160 X 22 X 20 mm, were solution annealed at 1090˚C for 1 hour followed by water quenching and further aged at 650˚C for 20000 hours [7, 16]. The samples for optical microscopy have been prepared by the conventional route of grinding and polishing followed by vibratory polishing as the final step. In vibratory polisher, both colloidal silica and alumina suspension have been used separately for revealing various microstructure features. Electron transparent thin samples for characterization using analytical transmission electron microscopy were prepared by twin jet polishing using perchloric acid : methanol (in 1:9 ratio) electrolyte. Diffraction contrast images and selected area diffraction patterns were recorded by transmission electron microscopy in Philips CM200 ATEM operated at 200 kV equipped with Energy Dispersive X-ray Spectroscopy (EDS) detector. HAADF- STEM (Scanning transmission electron microscopy by high angle annular dark field) imaging and EDS elemental mapping has been carried out in Talos (Thermofisher Scientific) at an accelerating voltage of 200 kV. JEMS electron microscopy simulation software version 4.46 [17] was used for the analysis and simulation of diffraction patterns.

*Table I. Chemical Composition (in wt %) of SS 316LN Austenitic Stainless Steel*

| Sample | Cr | Ni | Mo | Mn | C | P | S | N |
|---|---|---|---|---|---|---|---|---|
| N07 | 17.5 | 12.2 | 2.49 | 1.70 | 0.03 | 0.013 | 0.0055 | 0.07 |
| N11 | 17.5 | 12.2 | 2.5 | 1.72 | 0.03 | 0.013 | 0.0055 | 0.11 |
| N14 | 17.5 | 12.1 | 2.53 | 1.74 | 0.03 | 0.017 | 0.0041 | 0.14 |
| N22 | 17.5 | 12.3 | 2.54 | 1.70 | 0.03 | 0.018 | 0.0055 | 0.22 |

## 3. Results

Microstructural analyses of N07, N11, N14 and N22 steels have been carried out to study the morphological characteristics by means of optical microscopy and the chemistry and crystallography of precipitates using transmission electron microscopy. The observations and inferences are illustrated in the following sections.

*3.1. Morphological characterization of grains and precipitates using light optical microscopy*

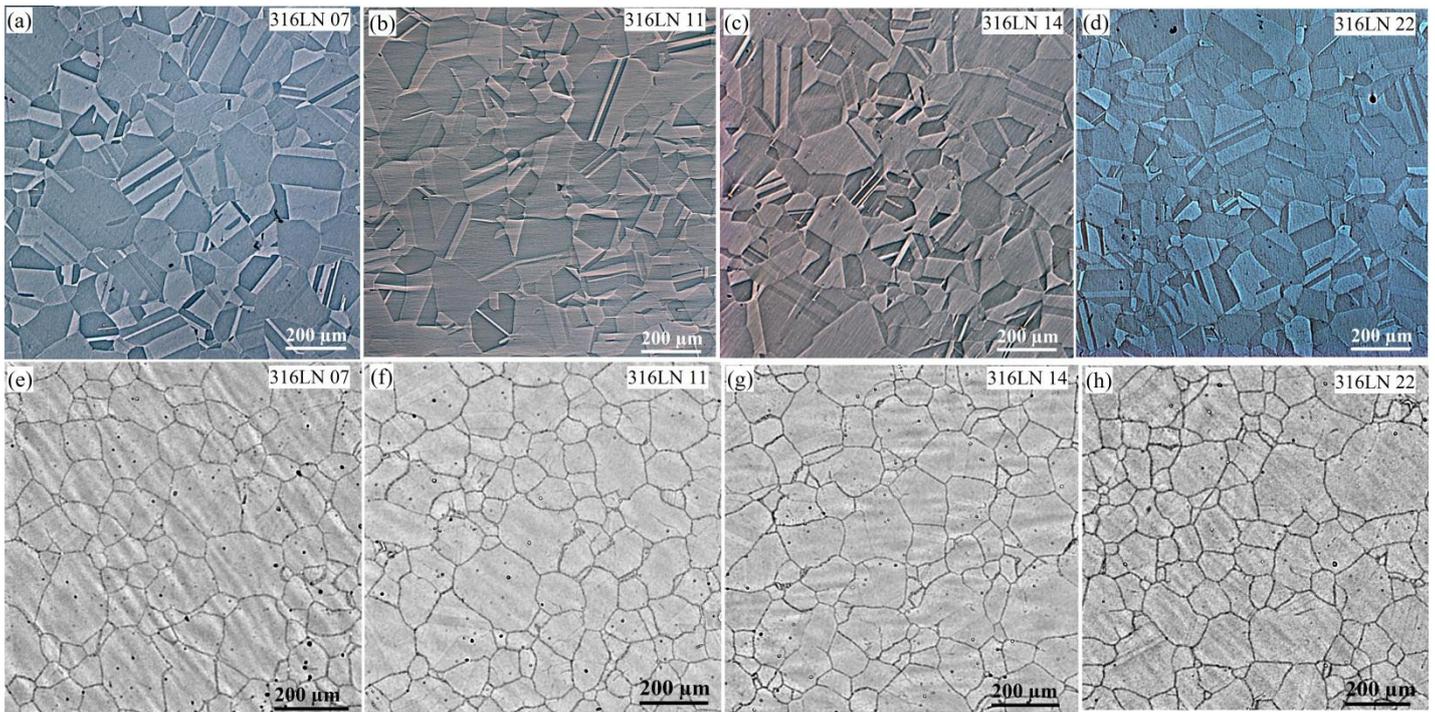

Figure 1. Optical micrographs of N07, N11, N14 and N22 showing the contrast of (a-d) annealing twins (final polishing with alumina suspension), and (e-h) nearly equiaxed grains with marginal bimodal division (final polishing with colloidal silica)

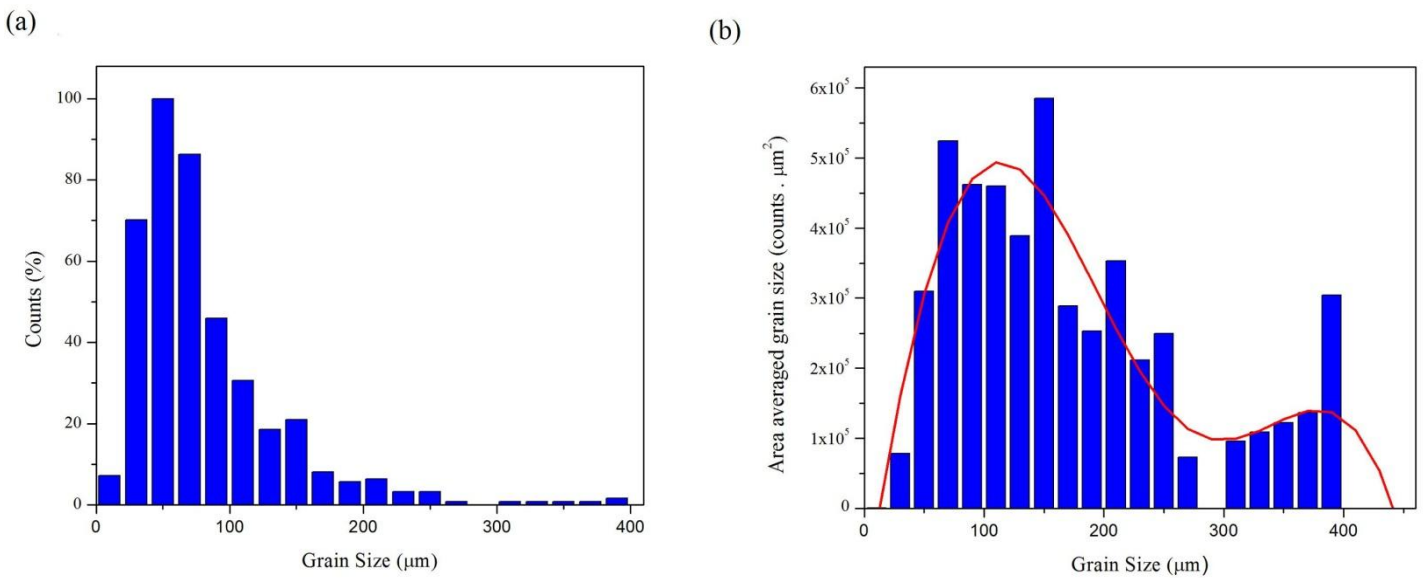

Figure 2. Graph shows grain size distribution analysis from N22 sample: (a) Normalized counts vs. grain size graph and (b) Area averaged grain size vs. grain size graph show the trend of bimodal division in grain size distribution.

Figure 1(a-d) depicts optical images from samples prepared by vibratory polishing using alumina suspension. The bright field optical contrast shows the presence of annealing twins. In homogenized SS 316 LN, the most favorable annealing twins lie on {111} planes. The contrast depicts the V-shaped double twins randomly distributed in several grains, which can be represented by the sides derived from the well-known concepts of the Thompson tetrahedron [18, 19]. The geometric relationships of four equivalent {111} slip planes in fcc metal are illustrated by the Thompson tetrahedron. The angle between any two slip planes is 70.53º (or 109.5 °) and which may rarely visible due to the off-beam projection of grains. Qualitative analysis

of these micrographs reveals the resemblance in twin morphology thereby implying that nitrogen concentration has a negligible role on twin density.

Optical microscopy reveals the grain microstructure (Figure 1(e-h)) in samples prepared by the second method where colloidal silica has been used as the final polishing agent in the vibratory polisher. Bright field contrast in optical imaging projects a ditch microstructure for the grains indicates the sequential formation of grain boundary precipitates which will be discussed in subsequent sections. The precipitate substructure emerges predominantly because of the CMP (chemical mechanical polishing) action of colloidal silica. Equi-axed grains of size ranges from 10 μm to 410 μm are visible in all samples with a slight bimodal division in morphology. By analyzing nearly 1000 grains from the N22 sample, the graph of grains size vs frequency counts (Figure 2a) has been drawn which shows the maximum counts at 50 μm and~ 350 μm for smaller and larger grains respectively. Figure 2b depicts the grain size vs. area average grain size where the heterogeneous grains with bimodal nature are demonstrated by Gaussian fit. The similarity of grain size distribution in each sample is obvious from the images. This analysis illustrates there is no significant variation in grain size and its distribution for the nitrogen concentration.

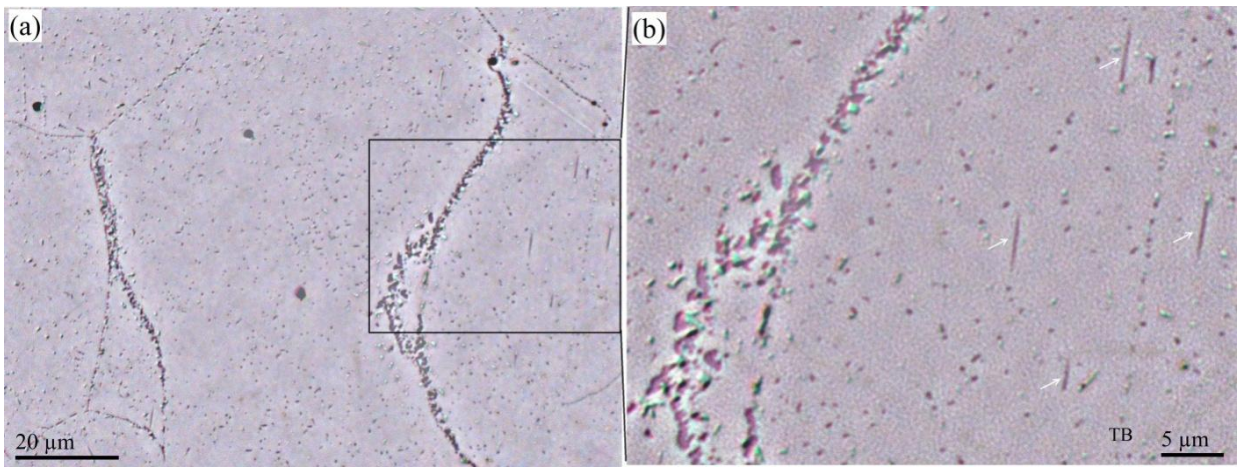

Figure 3. Optical images taken from N11 depict various precipitates by colour contrast (vibratory polishing with colloidal silica).

Figures 3a and b present the representative microstructure with much finer details of precipitate contrast. Figure 3b highlights a typical bright-field colour contrast at reflection mode in optical imaging (rectangle box) for various features/precipitates. While illuminating the sample, each region will alter the wavefront and hence the light absorption to give different optical disparities that emerge as unique colour contrast for each feature [20]. The contrast mechanism is similar to the shadow effect in etched samples, where the formation of reliefs and up/downs due to preferential polishing can happen differently with different polishing agents. Polishing parameters of vibratory polisher have a crucial role to play in projecting the required features in microstructure, thus optimization for polishing cloth, solution, and time has been carried out for each sample.

The colour contrast due to optical disparities depicts the presence of substantial amount of intra-granular and inter-granular precipitate formation with two major colour variations; amaranth and green. Lean grain boundaries are decorated by cascades of elongated amaranth precipitates (thickness ~ 0.5 μm, length ~ 1 – 2 μm) tangential to the boundary, whereas irregular bigger precipitates (size ~ 1 – 2 μm) are common in densely populated grain boundaries. The green features form adjacent to the amaranth ones in both thin and dense boundaries. The non-uniform distribution of inter-granular precipitates can be attributed to the degree of strain at the grain boundaries. It has been also observed that the twin boundaries are partially covered with precipitates, mostly amaranth ones. The intra-granular precipitates of both amaranth and green type are randomly distributed all over the grain. However, there is a considerable variation in the number density, size and morphology of the precipitates while considering the images from each sample (Figure 4) with varying nitrogen content. For example, some amaranth precipitates selectively elongate up to 5μm (indicated with white arrow in Figure 3b) in some samples, meanwhile, the green ones elongate arbitrarily. Moreover, an increase in total secondary phase formation with nitrogen concentration has been observed. Before commenting on each coloured feature, the chemistry and crystallography have been characterized by transmission electron microscopy as follows.

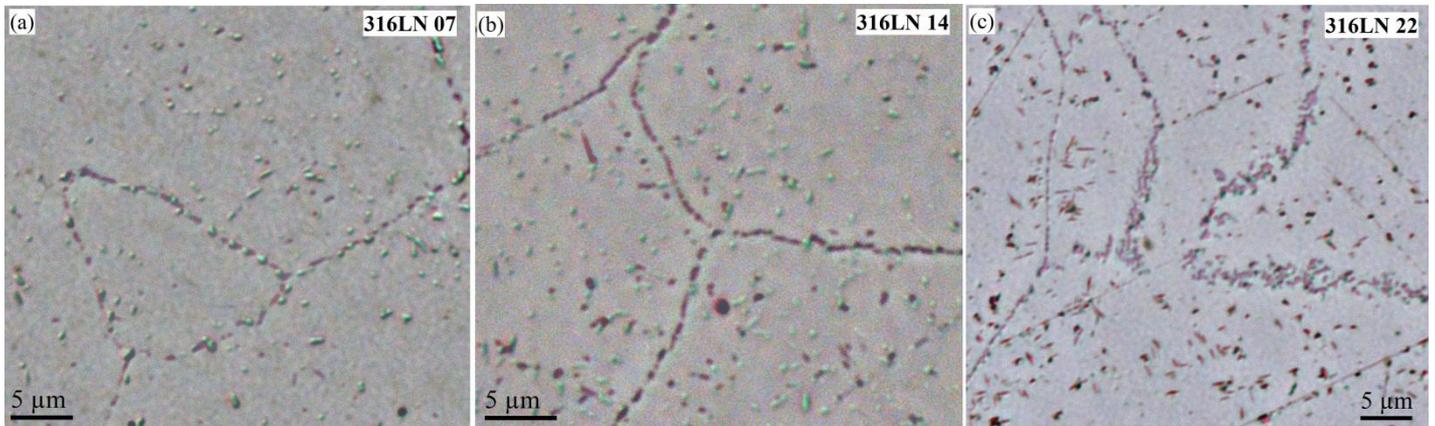

Figure 4. Optical micrographs depict precipitate sub structure of N07, N14 and N22 steels

*3.2. Identifying the crystallography of precipitates: analysis using transmission electron microscopy*

The TEM micrographs in Figure 5 demonstrate the evolution of precipitate microstructure in SS 316LN during long-term aging. The bright field image (Figure 5a) from sample N07 depicts the distribution of inter-granular and intra-granular precipitate phases. A figure from same region has been reported in our previous article [16], however characterization of precipitate contrasts have been completed in the present study including the simulation experiments. Precipitate phases have faceted cuboidal, elongated, and irregular morphologies with size ranges from 100 nm to 1μm. The SAD (selected area diffraction) pattern recorded from grain interior in Figure 5b and c shows fcc γ matrix along <112>γ and <-110>γ zone axis respectively, where reflections from cuboidal shaped precipitates also present. Overlapping of the third-order precipitate reflection with the first order matrix reflection implies the cube-on-cube orientation relationship between matrix and precipitate with coherency. These correlations are confirmed by simulating the diffraction patterns of γ matrix (a = 3.56 Å, Fm-3m) and $M_{23}C_6$ precipitate (a = 10.67 Å, Fm-3m) using JEMS (Java-based Electron Microscopy Software) [17] and by generating the combined diffraction pattern as shown in Figure 5d. Chemical analysis using TEM-EDS indicates the presence of Cr, Fe, and Mo in this carbide in descending order of composition. The point EDS experiment also reveals the presence of nitrogen along with carbon, thus the precipitate can be more specific as $M_{23}(C, N)_6$ carbo-nitrides.

The Cr-enriched $M_{23}(C, N)_6$ precipitates are also populated in grain boundaries but differ in size and shape (marked on Figure 5a). They are comparatively larger in size with irregular morphology. Diffraction analysis on these grain boundary carbides reveals that they have a coherency and cube-on-cube orientation relationship with one of the matrix grains from which it has been grown. On the coherent matrix side, the precipitate has a faceted morphology while shape irregularities are mostly visible on the other side.

Figure 6 shows the coherent $M_{23}(C, N)_6$ precipitates occupied in the twin boundary and near regions. By characterizing the selected area diffraction pattern taken from the twin boundary, the zone axis has been identified as $[011]_γ$ and the twin plane is $(111)_γ$, which is a coherent twin. The traces of coherent precipitates in both sides of the twin plane with a cube-on-cube orientation relationship are visible in diffraction patterns as satellite reflections. Figure 6b shows the dark field image taken from the selected reflections (which include the reflections from precipitates embedded in both sides of the twin) shown inside the white circle in the SADP. Thus the diffraction contrast corresponds to the cuboidal carbides of ~ 200 nm situated in both sides of the twin has been imaged. A faint contrast of one of the carbides with finer dimensions on the twin boundary grown towards one side of the twin is also visible in the image (white arrow). It is observed that, in the twin boundary, the coherent precipitates like $M_{23}(C, N)_6$ always grow towards either side of the twin but across the boundary.

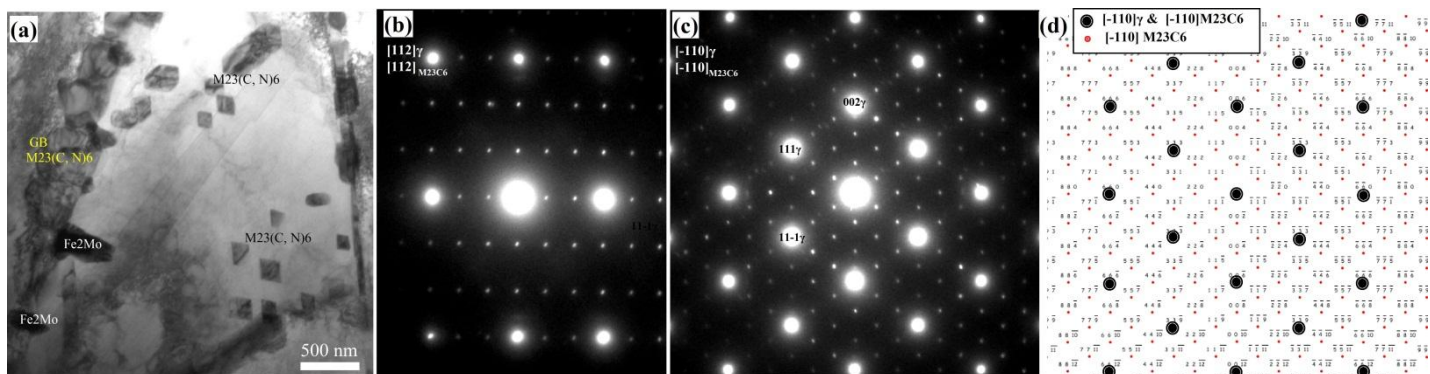

Figure 5. TEM micrographs shows typical microstructure emerged in SS 316LN (N07) after long-term aging (GB-Grain boundary). (a) Bright field image [16]; (b) SADP in 112γ; (c) SADP in 110γ (d) Simulated diffraction patterns of γ matrix (a = 3.56 Å, Fm-3m) and $M_{23}(C, N)_6$ precipitate (a = 10.67 Å, Fm-3m) combined.

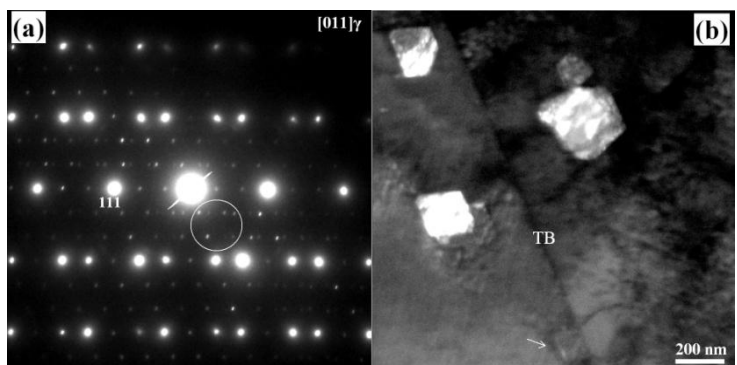

Figure 6. TEM characterization on TB (twin boundary): (a) SADP recorded on [011] zone axis and (b) Dark field image corresponds to the precipitate reflections inside the circle in Figure 6a.

Figure 7 shows the bright field image, selected area diffraction pattern and dark field image recorded from the second kind of precipitate found in all steels. The chemical analysis using EDS shows that the precipitate is rich in elements such as Fe and Mo. The diffraction pattern analysis (Figure 7b and d) shows the structural dimensions of this phase match with η-$Fe_2Mo$ Laves Phase having hexagonal symmetry. The diffuse streaks along the preferred direction in the diffraction pattern and lamellar morphology in bright field and dark field images indicate the presence of planar defects/stacking faults in the intermetallic phase. These elongated precipitates with crammed planar defects are also found nearby the grain boundaries (Marked as $Fe_2Mo$ in Figure 5a).

Furthermore, in high nitrogen steels, N14 and N22, it has been observed that $Fe_2Mo$ intermetallic are mostly contiguous with yet another precipitate phase as seen in Figures 8 and 9. Diffraction analysis with simulation experiments in Figure 9a and b [16] confirms these are $Cr_2N$ precipitates with crystallographic features of hexagonal symmetry of Pearson's symbol hP11. The diffraction contrast images clearly demonstrate the coexistence of $Cr_2N$ with $Fe_2Mo$ intermetallic phase. It is highly likely that the Laves phase crystals act as heterogeneous nucleation sites for $Cr_2N$ precipitate. Scanning transmission electron microscopy by high angle annular dark field imaging (STEM-HAADF) combined with EDS mapping (Figure 10) gives an unambiguous depiction of chemical composition in each phase. Substantial nickel enhancement is observed in intermetallic, however other substitutional elements are completely absent. Moreover, manganese and carbon are preferentially segregated within $Cr_2N$ precipitates, by replacing chromium and nitrogen atoms correspondingly.

From selected area diffraction pattern analysis of varieties of precipitates following conclusions can be made: the grain boundary precipitates are carbo-nitrides particularly $M_{23}(C, N)_6$ precipitates and $Fe_2Mo$ intermetallic form adjacent to that. The twin boundaries are decorated mostly with $M_{23}(C, N)_6$ precipitates with faceted cuboidal morphology. Inside the grain, a substantial amount of cuboidal $M_{23}(C, N)_6$ precipitates and elongated/irregular $Fe_2Mo$ intermetallics are randomly distributed. In high nitrogen steels, N14 and N22, $Cr_2N$ forms frequently with $Fe_2Mo$ intermetallic by sharing the interface in a contiguous manner.

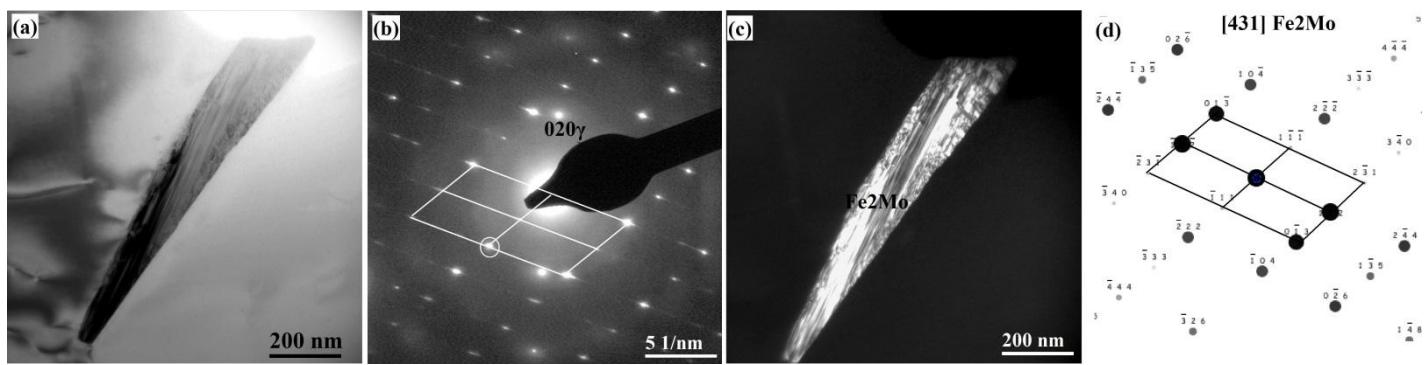

Figure 7. TEM characterization on $Fe_2Mo$ intermetallic: (a) Bright field TEM image; (b) SADP taken from the intermetallic shows the traces of planar defects as streaking in preferred direction; (c) Dark field image of intermetallic taken from circled reflection in Figure 7b, and (d) Simulated diffraction pattern on [431] zone axis of $Fe_2Mo$ matches with the experimental pattern in Figure7b.

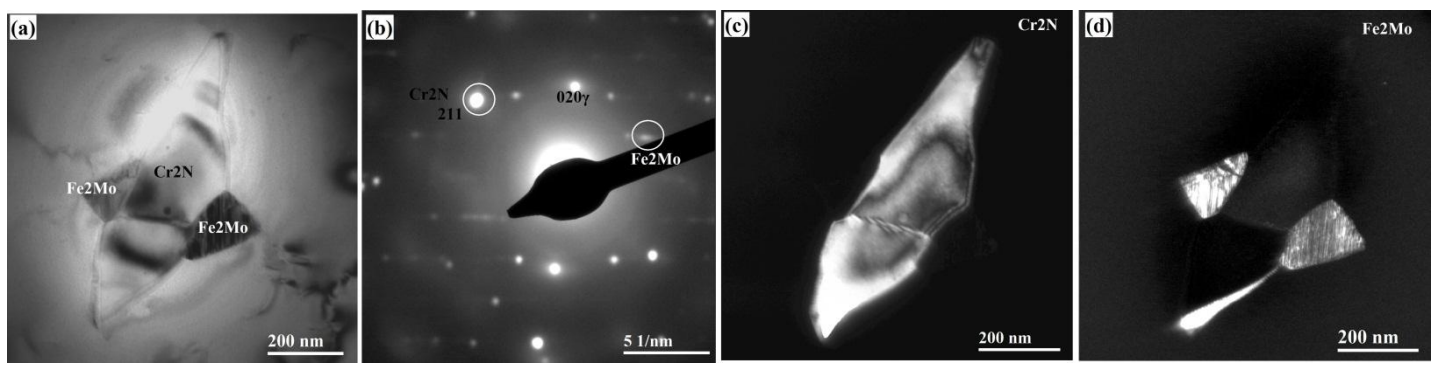

Figure 8. TEM characterization of contiguously formed $Cr_2N$ with $Fe_2Mo$ intermetallic in N14: (a) bright field image; (b) selected area diffraction pattern from precipitates; (c) dark field image from $Cr_2N$; (d) ) dark field image from $Fe_2Mo$.

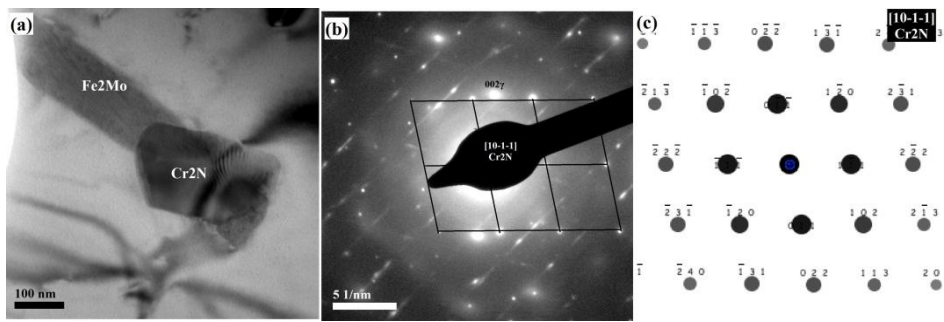

Figure 9. TEM characterization of contiguously formed $Cr_2N$ with $Fe_2Mo$ intermetallic in N22: (a) Bright

field TEM image [16]; (b) Selected area diffraction pattern from precipitates [16]; (c) Simulated diffraction pattern of $Cr_2N$ in [10-1-1] zone axis.

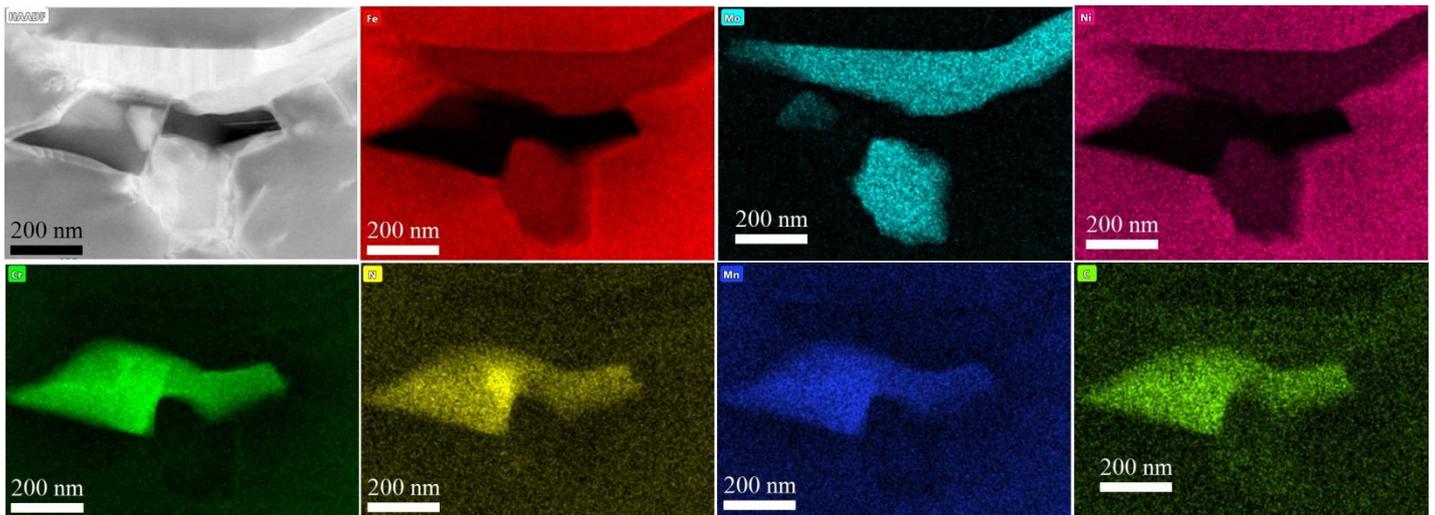

Figure 10. Scanning transmission electron microscopy by high angle annular dark field imaging (STEM-HAADF) combined with Energy dispersive X-ray (EDS) mapping.

By comparing with transmission electron microscopy studies, the phase symmetry of each precipitate/feature with colour contrast in optical micrographs has been identified. The amaranth features found in grain boundaries and inside grains are carbo-nitrides, $M_{23}(C, N)_6$ precipitates. The randomly elongated or irregular green features form near the grain boundaries and inside the grains are $Fe_2Mo$ intermetallics. A thorough investigation in optical micrographs unveils the picturisation of contiguous precipitation of $Fe_2Mo$ and $Cr_2N$ as seen in Figure 11, where typical red feature is frequently associated with green ones (also seen in Figure 4b-c). The contrast of this 'red' colour (represents $Cr_2N$) is slightly different from 'amaranth' ($M_{23}C_6$) and can be easily recognized as they are always being adjacent to the 'green' ones ($Fe_2Mo$). The contiguous $Cr_2N$/ $Fe_2Mo$ formation is mainly in high nitrogen steels, N14 and N22 steels, which increases with nitrogen content. However, a highly isolated $Cr_2N$/ $Fe_2Mo$ combination has been anticipated from optical images from N07 and N11 samples by analyzing the topographic feature, exposed by colour contrast. Since optical microscopy gives a broad view of these categorized precipitates, the frequency of occurrence, morphologic derivatives can be easily visualized from images. For example, Figure 11 projects the various morphological entities of contiguous precipitates as (a) $Fe_2Mo$ side by side with $Cr_2N$ (Green / Red combination), (b) $Cr_2N$ sandwiched between two $Fe_2Mo$ (Green/ Red/ Green combination), (c) $Fe_2Mo$ sandwiched between two $Cr_2N$ (Red/Green/ Red combination), and (d) more than 2 crystallites of each precipitate sandwiched alternately. The latter one is very frequent in N22 steel as seen in Figure 11b.

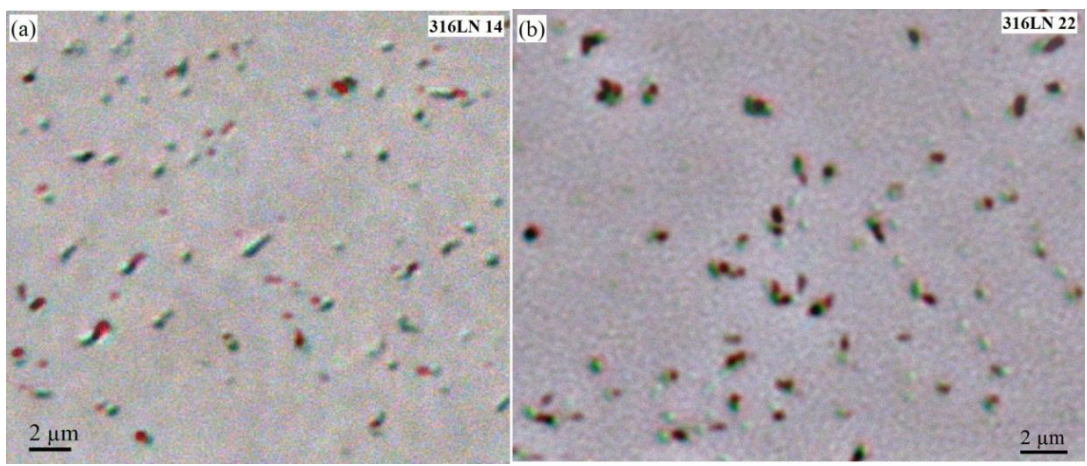

Figure 11.Colour contrast in optical micrographs shows the contiguous precipitation of $Cr_2N$ with $Fe_2Mo$ as joined red and green features in (a) 316LN 14 and (b) 316LN 22.

*3.3. Preferentially grown $M_{23}(C, N)_6$ stringer along $(111)\gamma/[110]\gamma$*

Optical micrographs provide an expansive view of precipitate distribution and geometry easily and precisely, thus, an interesting phenomenon of stringer morphology for $M_{23}(C, N)_6$ precipitates (marked in Figure 3b and 12) has been observed inside the grains. Micrographs prove these stringers form prominently in N07, N11 and N22 steels. In N14 steel (Figure 4b and 11a) the stringers are hardly seen. In N22 steel, stringers are heavily populated in such a way that the majority of amaranth precipitates/$M_{23}(C, N)_6$ are elongated (Figure 4c and 12c). The maximum length of the stringer can be measured as 5 µm if the long axis of the stringer is parallel to the planar view of the image (stringers in Figure 12c from N22 steel appear shorter because of the off-axis view), and width is nearly 500nm. Observations demonstrate that the growth direction of stringers is preferential consequently, there are multiple crystallographic variants exist within single grain. In Figure 12a, the stringers from the same grain indicated by white arrow lie along 10'o clock position whereas, that by black arrow in 2'o clock position, which illustrates the presence of two variant selections. Thorough investigation in optical images reveals, generally, the stringers are parallel to the twin as seen in Figure 12b, which will provide a hint to identify the directional preferences of the precipitate growth. While relating with the well-known fact that, the annealing twins usually form on close-packed {111}/<112> plane/direction and are coherent for fcc structure [21-23], it can be confirmed that the stringers of $M_{23}(C, N)_6$ grow on {111} planes of austenite matrix.

To find the growth direction of $M_{23}(C, N)_6$ stringer on $\{111\}_\gamma$ plane, two major possibilities has to be analyzed: (a) twin direction of {111} plane i.e., $<112>\gamma$, and (b) slip direction of {111} plane i.e., $<110>\gamma$. The analysis is based on the first principle models predicting the ORs (orientation relationships) and the corresponding habit plane with matrix [24-27]. Accordingly, the criteria for the formation of low energy interface between two phases are: (a) have fully or partially coherent orientation relationship which reforms morphology, and (b) the matching atom rows should be close packed or nearly close packed [28]. In the present scenario of phases with cubic symmetry, γ matrix ($a_\gamma$ = 3.56 Å, Fm-3m) and $M_{23}(C, N)_6$ precipitate (a = $3a_\gamma$=10.67 Å, Fm-3m) have cube-on-cube orientation relationship where all planes and directions are parallel, hence the conditions are perfectly accomplished. As per the first criteria, morphology/growth directions of precipitates are consistent with the degree of mismatch at precipitate-matrix interfaces [29]. Minimum mismatch, hence favorable interfacial energy is attained by the parallelism between the densely packed planes and directions, i.e., $\{111\}/<110>\gamma$. For fcc, the twin direction, $<112>\gamma$ is the least populated direction and is not favored for the growth of precipitates. Thus the preferred growth plane and direction for $M_{23}(C, N)_6$ stringer can be deduced from optical micrographs as {111}/<110>, where six crystallographic variants are allowed.

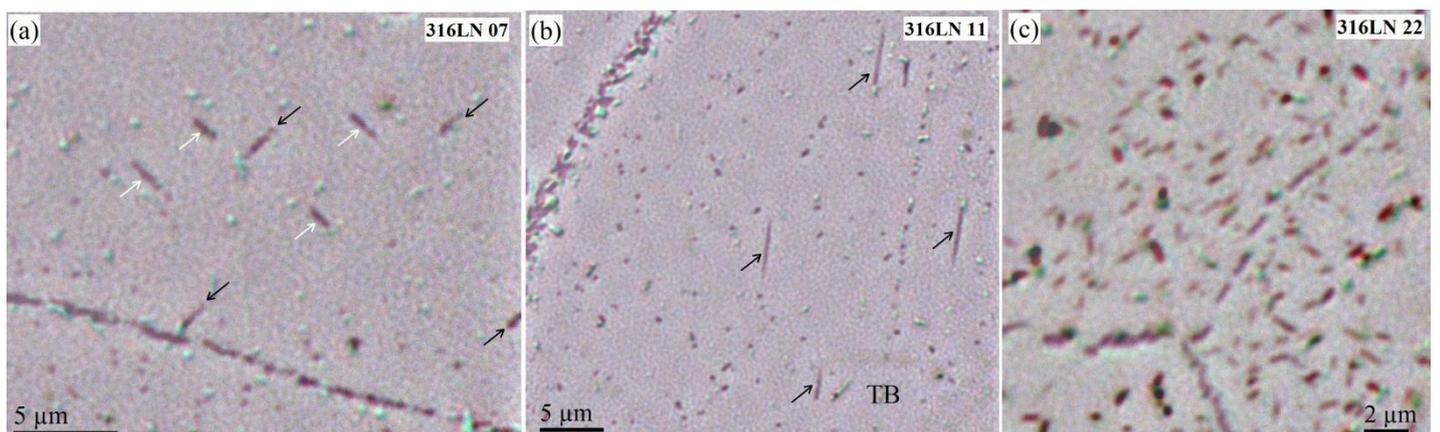

Figure 12. Primary stringers of $M_{23}(C, N)_6$: Optical micrographs of preferentially grown $M_{23}(C, N)_6$ stringers

having colour contrast of amaranth (lamellar) appear in (a) N07, (b) N11 and (c) N22 samples.

*3.4. Clustering of preferentially grown $M_{23}(C, N)_6$ stringers*

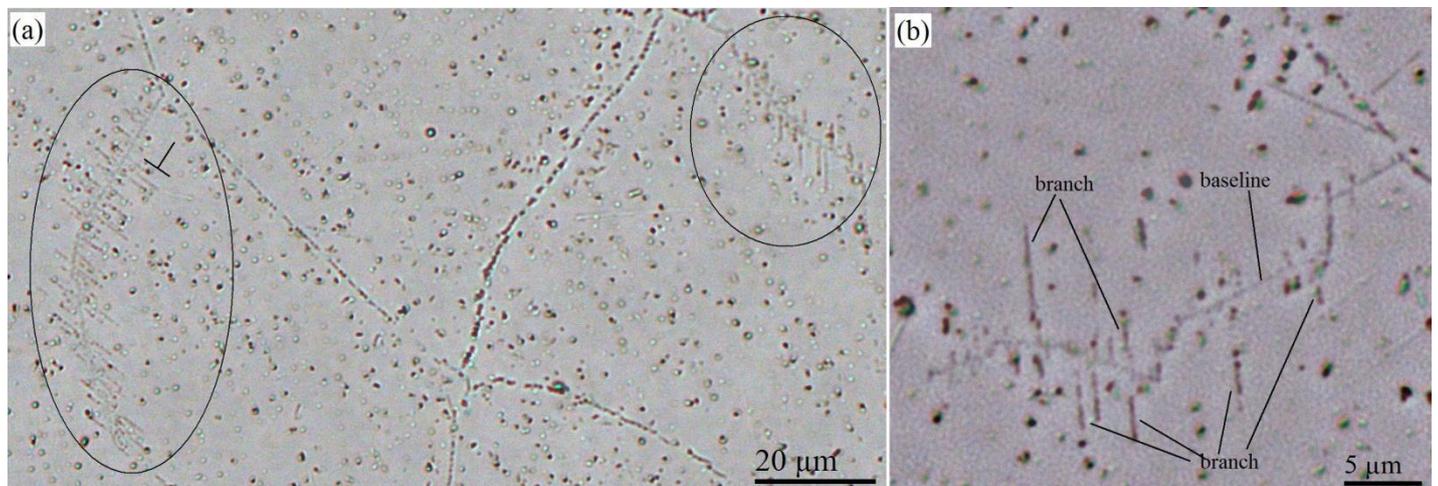

Figure 13. Secondary stringers of $M_{23}(C, N)_6$: Optical micrographs from N22 samples prepared in vibratory polisher with silica colloidal solution shows distinct features: (a) Branching of amaranth precipitates/$M_{23}(C, N)_6$ on a common baseline in a preferential direction, (b) The highlighted image shows a better contrast for the textured precipitate as clusters of stringers/branches in a common baseline.

In addition to the stringer morphology, a very unique formation process of $M_{23}(C, N)_6$ precipitate has been identified in N22 steel with high nitrogen content as depicted in Figure 13. Figure 13a shows a typical sequential formation of precipitates (circled), where the preferentially grown branches/stringers sharing a common baseline. Each branch of ~10 μm length forms parallel in a common habit plane within ~ 1 μm interval. Generally, these clusters are grown up to hundreds of micrometers. Typical features observed in these images are (1) the baseline may shift or bend by its trajectory, however, the branches keep parallelism consistently, (2) there are regions where the baseline plane is perpendicular to the growth direction of branches (Marked in Figure 12a), (3) preferentially grown $M_{23}(C, N)_6$ stringers (Figure 12) are comparatively less in the vicinity of the clusters of stringers, and (4) the ends of the baseline of each cluster can be a grain boundary or a twin boundary or mixed.

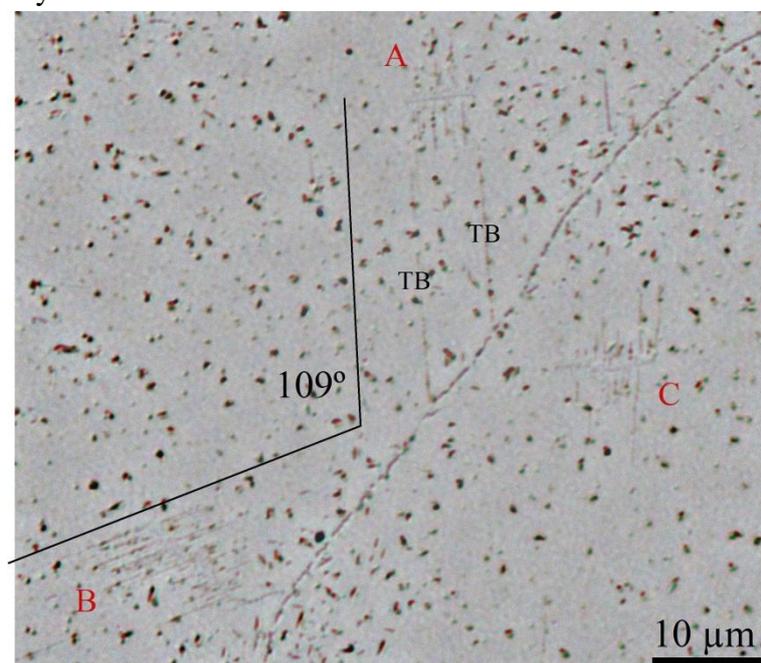

Figure 14. Secondary stringers of $M_{23}(C, N)_6$: Two variants of clusters of preferentially grown $M_{23}(C,$

N)$_6$ stringers are identified in single grain indicated as 'A' and 'B', where A starts and ends at twin boundaries (TB).

In Figure 14, three clusters of preferentially grown stringers are evident (marked as A,B and C), where 'A' and 'B' are from the same grain, whereas 'C' forms nearby grain. Close observation in clusters reveals the morphological characteristics of each. Cluster 'A' starts and ends in twin boundaries, whereas, 'B' and 'C' originate in grain boundary but the other end diminishes in the matrix. Even though there is poor contrast at the end of its baseline, a detailed analysis throughout the sample hints the role of twin boundary in terminating the cluster formation.

Similar to the previous section (section 3.3), the directional correspondence of these clusters of preferentially grown $M_{23}(C, N)_6$ stringers can be inferred exclusively from optical micrographs by relating with the twin planes. In cluster 'A' the branches/ stringers are formed on the planes parallel to twins on $\{111\}\gamma$ planes, indicating $\{111\}$ plane is the growth plane. In the same grain there is another cluster 'B' where the branches are grown along 109º angle from the plane of branches in 'A'. The angle between two opposite body diagonal (along <111>) is 109.5º and which can be measured exactly in the zone axis of <110>$\gamma$. Hence the grain with 'A' and 'B' clusters lie exactly at <110>$\gamma$ zone axis. As it has been discussed in the previous section, the growth direction of precipitate governs by dense direction, which is <110>$\gamma$ for $\{111\}$ plane. In this context, the possibilities of the formation of incoherent twins on $\{112\}$ planes cannot be eradicated, as the angle between these planes also 109.5º. However, out of two favorable directions, <110> and <111>, <110>$\gamma$ is the most densely populated one for $\{112\}$ direction as well. Thus the stringers' growth direction can be confirmed as <110>$\gamma$. In nutshell, the image depicts two different clusters of preferentially grown $M_{23}(C, N)_6$ stringers in a different set of $\{111\}$ planes in single grain in the viewing direction of <110>$\gamma$, which proves the presence of variants even for cluster formation.

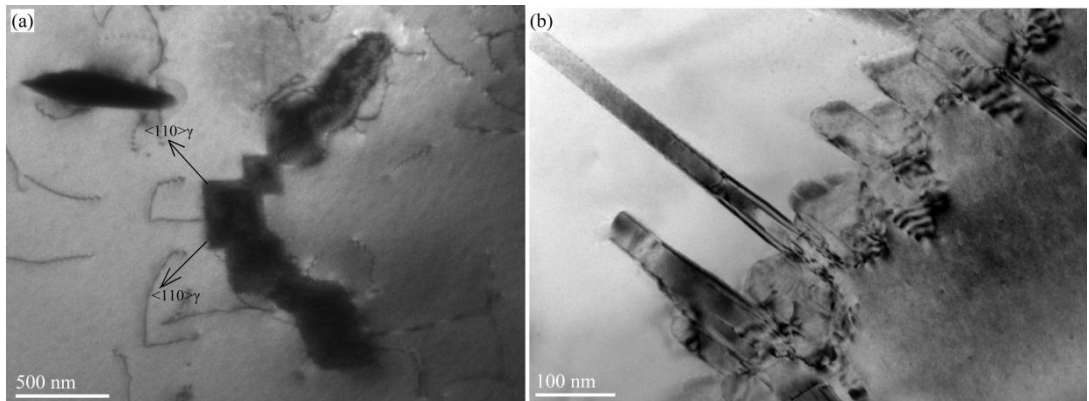

Figure 15. Secondary stringers and lath clusters of $M_{23}(C, N)_6$: Diffraction contrast TEM images from N22 resolves two kinds of clusters of preferentially grown $M_{23}(C, N)_6$. Cluster formation of (a) strings of cubes/stringers of $M_{23}(C, N)_6$ (b) laths of $M_{23}(C, N)_6$.

Further, the evidence for directional preferences has been demonstrated by transmission electron microscopy, in addition to the optical microscopy findings. Nevertheless, it is tricky to get imaged a precipitate stringer of 5 μm length with preferential growth from an electron transparent film of <100nm thickness. It is possible if the grains with precipitates are aligned perpendicular to <110>$_\gamma$ direction, the viewing direction. Figure 15 shows the TEM bright filed images where the diversity in branching has emerged as strings of cubes (Figure 15a) and clusters of laths (Figure 15b) which provide further confirmation on crystallinity, orientation relationship and preferential growth direction. The figure establishes the morphology of stringers are strings of cubes of $M_{23}(C, N)_6$. These two different morphologies for clusters of preferentially grown $M_{23}(C, N)_6$ stringers could be resolved by doing diffraction contrast imaging in TEM. Figure 15a shows a part of cluster of $M_{23}C_6$ stringers and the direction of growth has been proven as <110> direction by diffraction analysis. This can be also confirmed from the shape of these precipitates seen in diffraction contrast images, which are perfectly cuboidal with cube- on –cube orientation relationship with the matrix. The image is projected along nearly $[001]_\gamma$ direction and $<110>_{\gamma/M23C6}$ direction can be easily distinguished by its cubical geometry. Figure 15b establishes the second morphological derivative for clusters of preferentially grown $M_{23}(C, N)_6$ with lath structure. Each lath is packed on a common strand and is grown towards $<110>\gamma$ direction with cube- on – cube orientation relationship with the matrix.

## 4. Discussion

*4.1. Colour contrast for precipitate phases*

In literature, there are numerous studies that distinguish different precipitate phases in optical micrographs by chemical etching or colour tint etching [30, 31]. However, in the present study, vibratory polishing at optimizing conditions with colloidal silica and alumina suspensions for 24 hours has been implemented for revealing the micro-structural features without any further etching process. The twins have been emerged by bright field reflection mode in optical microscopy by using alumina suspension whereas precipitate phases have been distinguished by using colloidal silica as the final polishing agent. Optical lithography takes place by mechanical action with nanometer-sized abrasive particles of alumina which enables visualization of twin microstructure. Both mechanical and chemical action (CMP-Chemical mechanical polishing) from the solution additives in colloidal silica [32] that causes preferential material removal gives the precipitate substructure. Amaranth, green, red and grey colours represent $M_{23}(C, N)_6$, $Fe_2Mo$, $Cr_2N$ and matrix respectively. Microscope optics including the illuminating system and the imaging system is capable of giving the contrast from distinct optical disparities. These disparities are materialized from the light absorption and refractive index of the material, which is intrinsic for each precipitate. Add on, the formation of up/downs on the surface due to the hardness variation of precipitate phases impart a shadow effect causing the contrast in the image. Thus the colour contrast assigned to each precipitate phase is the combined effect from the vibratory polishing techniques and optics of the image formation in bright field optical imaging.

*4.2. Bimodal nature in grain size distribution of SS 316LN*

All steel varieties with varying nitrogen content in the present study exhibit a similar trend of bimodal distribution with the coarse and fine grains. The difference is narrow between the maxima of smaller grains (50 µm), and bigger grains (350 µm) with nearly single order magnitude. However, the effect of grain distribution in overall mechanical properties and engineering performances will be significant because bimodal grain distribution is one of the microstructural strategies for implementing a good combination of strength and ductility to the system. Literature reports a bimodal division (20 - 250 µm) for SS 316LN with nitrogen concentration of 0.174 wt. % [33], where it is shown that the range of grain size variation will be minimal for lower nitrogen concentration. Bimodal nature of grain distribution has been observed in the solution annealed 316LN as well. Images from existing literature show the proof [34, 35] however, the observation has not been specifically mentioned. Recently, studies on nitrogen added High Entropy Alloys (HEA) illustrate the heterogeneous grain size distribution trending to have a bimodal division [36] which won't compromise the strength-ductility trade off. Bimodal division of single order of magnitude of SS 316LN may contribute to the superior creep resistance property along with other upgraded mechanical properties. However, the current study doesn't observe any significant variation of the bimodal trend with nitrogen content for a limited range of nitrogen concentration from 0.07 to 0.22 wt. %.

*4.3. Contiguous hetero-structures of $Fe_2Mo/Cr_2N$ in SS 316LN*

Literature reports the possibilities of formation of σ phase and χ phase in high temperature aging (between 850 ˚C and 900 ˚C) theoretically and experimentally [37]. $Cr_2N$ precipitation in 17% Cr, 13% Ni, 5% Mo low carbon steel with 0.25 wt% nitrogen is first reported in 1969 where $Fe_2Mo$ precipitation also has been identified [38]. The formation of $Fe_2Mo$ intermetallic on $Cr_2N$/Matrix interface has been reported in 9Cr-

1Mo steels by analyzing carbon replica samples in transmission electron microscopy [39-41]. Co-precipitation of $Fe_2Mo$ and $Cr_2N$ by contiguous way has been identified by amplitude contrast imaging in high alloyed austenitic stainless steel containing 0.54 wt.% of nitrogen [42]. The streaks in the SADP have been explained as the planar fault in the intermetallic. In contradictory, the joint formation of $Cr_2N/\sigma$ phase has been reported in steels containing elemental composition of 17.94 Cr; 18.60 Mn; 2.09 Mo; 0.89 N; 0.04 C; balance Fe in wt. % [43]. It has been observed that alloys with high manganese content are prone to have the side by side formation of $Cr_2N/\sigma$ phase [44-46].

In the present study, the analysis on contiguous $Fe_2Mo/Cr_2N$ (green/ red) precipitates quantifies the area density and frequency of occurrence from the optical micrographs with varying nitrogen content. By imaging large area, the average distance between two precipitate combinations has been roughlycalculated as 80 µm and 50 µm for N07 and N11steels respectively. In N14 steel, theoccurrence of contiguous precipitates repeats at 3 – 10 µm, whereas, in N22, the frequency is quite high;the co-existing $Fe_2Mo/Cr_2N$ of 1-2 µm size appears within 2 to 5 µm distance. The area density of $Fe_2Mo/Cr_2N$ precipitates has been measured as 0.0001 %, 0.005 %, 0.34 % and 2.7 % for N07, N11, N14 and N22 respectively. The colour contrast in optical micrographs enables to identify the least frequent occurrence of contiguous $Fe_2Mo/Cr_2N$ even in N07, the steel with the lowest nitrogen concentration in current study.

Yet another significant observation from the study is that, the absence of standalone $Cr_2N$ precipitates in all steels. $Fe_2Mo$ formation is definite in all samples and wherever the interface is compositionally favorable to form $Cr_2N$ it take place. Thus, with nitrogen concentration, the probability of favorable interfaces of $Fe_2Mo$/Matrix for the formation of contiguous $Fe_2Mo/Cr_2N$ increases and hence the number density. Prolonged aging aids chemical enhancement in both intermetallics and $Cr_2N$ by adding nickel and manganese/carbon respectively. The stoichiometry of the precipitates can be represented as $(Fe, Ni)_2Mo/(Cr, Mn)_2(N, C)$.

*4.4.Morphological derivatives of $M_{23}(C, N)_6$precipitates in steels*

SS 316LN subjected to long term aging shows abundance in $M_{23}(C, N)_6$ type carbo-nitrides in grain boundaries and inside the grains. The peculiarities of grain boundary carbides like size, shape, morphologyand crystallographic orientation depends on the nature of grain boundaries. Along low energy grain boundaries, the precipitate's growth is tangential to the boundary, elongated with the size of ~ 1µ length and 200-500 nm thickness. These carbides shows coherency with one of the grains on either side and have cube-on cube orientation relationship which enables bestpossible registry in order to minimize the interfacial energy. Each individual $M_{23}C_6$ carbide exists with similar orientation relationship with the matrix [47]. In contrast, carbides formed in high angle boundaries are big as ~1-2 µm and irregular in shape. High strain field in the boundary causes loss in coherency of precipitates with nearby grains. In twin boundaries cube shaped carbides of ~200 nm forms on either side of the boundary with cube on cube orientation relationship with the matrix.

While considering the intragranular $M_{23}(C, N)_6$ precipitates, four varieties of precipitate morphology has been observed as shown in the schematic diagram (Figure 16); (1) isolated cube- shaped precipitates dispersed all over the matrix, (2) Primary stringers: also called strings of cubes or string formation by edge to edge contact of cube- shaped precipitates preferentially grown in {111}/<110>, (3) Secondary stringers: clusters of preferentially grown branched precipitatesalong {111}/<110> from a common strand and (4) clusters of preferentially grown laths along {111}/<110>. There are three preferential directions in {111} plane may suggest the probability of three-dimensional formation of secondary stringers and lath clusters in the system. However, the picturisation is challenging by 2D imaging techniques implemented in the present study. Above all varieties of carbo-nitrides have cube on cube orientation with the matrix γ, hence the coherency assured, which is a well-established phenomenon in 300 series steels with high resolution phase contrast microscopy [48]. Isolated cube-shaped precipitates of size ranges from 200 to 300 nmare uniformly dispersed in N07, N11 and N14, whereas those are less in N22, where string formation is prominent.

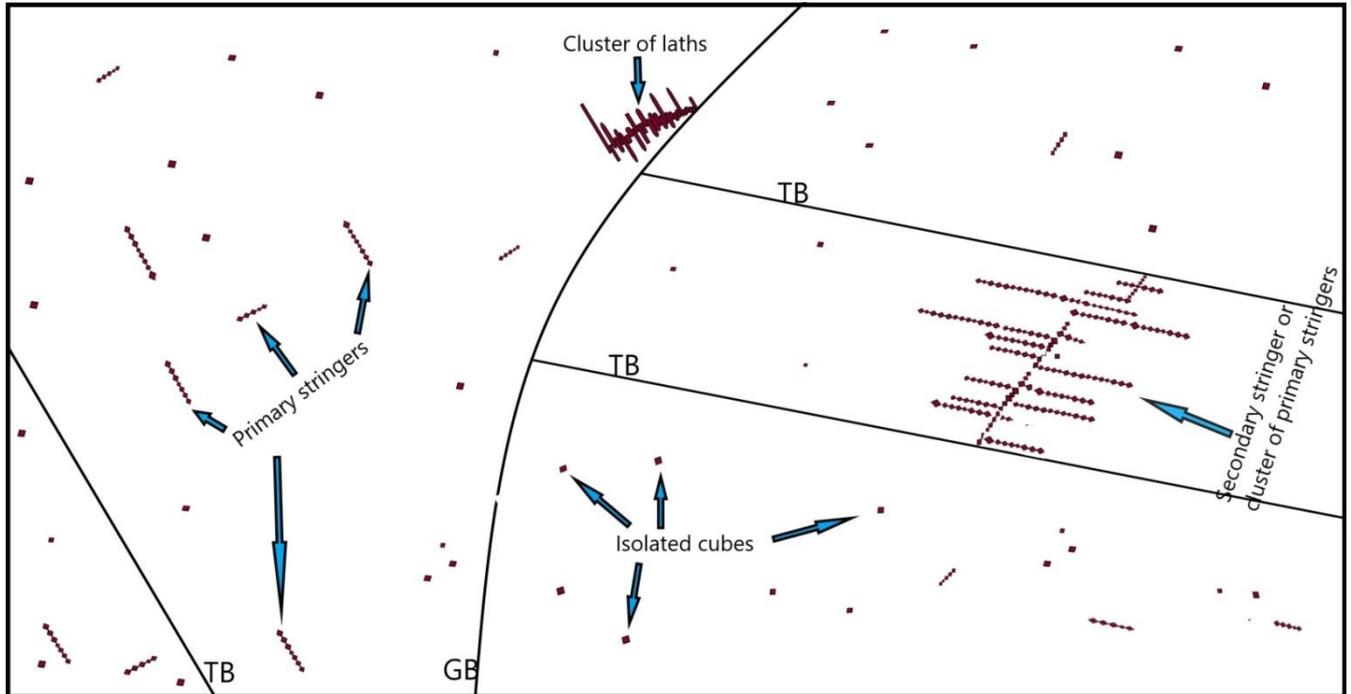

Figure 16. Schematic diagram shows four derivatives of intragranular $M_{23}(C, N)_6$ precipitates. ; (1) isolated cube- shaped precipitates, (2) Primary stringers : string formation by edge to edge contact of cube- shaped precipitates preferentially grown in {111}/<110>, (3) Secondary stringers: clusters of preferentially grown primary stringers as branches along {111}/<110> from a common strand and (4) clusters of preferentially grown laths along {111}/<110>. Crystallographic variants have been identified experimentally for primary stringers and secondary stringers.

The primary stringers are present in N07, N11 and N22 samples,and are lie in {111} planes along <110> direction. Crystallographic analysis along with light optical micrographs confirms the presence of six variants of primary stringers on <110> directions by imaging minimum of 2 variants in each grain. The number density measurements from 2-dimentional image of preferentially grown variants may give only a fraction of actual amount. However, the comparison of amount of stringers between samples is possible qualitatively as there is significant difference within each sample. Figure 4c and 11cdemonstrate that, the most of the intra-grain amaranth precipitates are elongated in N22where as in N07, N11 and N14 samples isolated cube shaped precipitates are prominent along with occasionally formed stringers which implies for higher nitrogen concentration, $M_{23}(C, N)_6$ is more prone to grow as stringers. Also it has been noted that, higher number density of stringers takes place in nearby regions of boundaries of grains or twins. In N14 steel, stringers are hardly seen, however, the formation of contiguous precipitation of $Fe_2Mo/Cr_2N$ is higher indicates the preference of chromium composition for switching from $M_{23}(C, N)_6$ stringer to $Cr_2N$ precipitates.

The presence of bothprimary stringers and secondary stringers of $M_{23}C_6$ has been reported in aged high alloyed high carbon (0.1- 0.2 %) steelsby using transmission electron microscopy [49-51]. Straight and intermittently curved primary stringers of length 10 – 20 µm has been visible occasionally in the images. The authors have predicted the growth direction of stringer lie along <112>γ from the trace analysis of stereographic projection. However, in the present study, it has been proved experimentally that, the dense plane and direction, {111}/<110> is the most favorable growth plane and direction. Additionally, theoretically it has proved in literature as explained in section 3.3 and 3.4 with the first principle calculations.

Cluster of preferentially grown laths has been described in literature by analyzing optical images as well as TEM images as continuous precipitates parallel to coherent boundary planes like twin boundaries [49]. It has been reported that continuous sheet or thick plates are grown parallel to {111} planes. As per literature yet another lamella of $M_{23}C_6$ grown on {111} planes nucleated on non-coherent twin boundaries [49, 50, 52]. In various alloy systems other than steels shows the abundance of formation of stringer morphology for $M_{23}C_6$ including Alloy 617 near to the grain boundary regions [53]. Thus the study establishes the effect of excess concentration of interstitial strengtheners in modifying the microstructural features in steels undergone long term aging.

*4.5. Precipitate sub-structure emerged with nitrogen content tailoring mechanical properties*

Following the identification of precipitates by unique colour contrast, quantification of each precipitate has been carried out by selecting each colour component by image processing as given in Table 2. Area fraction of precipitates in N07 is 1.6 % which increases substantially up to 12.5 % for N22 steels. The percentage of area fraction for contiguous $Fe_2Mo/Cr_2N$ is negligible for N07 and N11, but for N14 and N22 the values are hyped to 0.34 % and 2.7 % respectively. It is interesting to realize that N14 steel shows an abrupt variation in contiguous precipitation while comparing with N11. In N22 most of the $Fe_2Mo$ (green) precipitates are in contiguous with $Cr_2N$.

*Table 2. Quantification of precipitates in SS 316LN*

| Sample | Area fraction (%) | |
|---|---|---|
| | Total precipitates | Contiguous $Fe_2Mo/Cr_2N$ |
| N07 | 1.6 | 0.0001 |
| N11 | 2.7 | 0.005 |
| N14 | 5.2 | 0.34 |
| N22 | 12.5 | 2.7 |

Apart from change in total precipitate density, the evolution of derivatives of intra-granular $M_{23}(C, N)_6$ precipitates vary with nitrogen content (summarized in Table 3). Isolated cubes, common in all steels line up and transform into primary stringers in N07, N11 and N22 steels. Adding excess nitrogen introduces clusters of stringers and laths, as seen in N22. The exceptionality shown by N14 is the absence of primary stringers, even though it occurs in lower concentrations, implies the composition is energetically favorable for the formation of contiguous $Fe_2Mo/Cr_2N$ than $M_{23}(C, N)_6$ stringers. A similar trend has been seen in aged duplex stainless steel, where growth of $M_{23}C_6$ ceases and that of $Cr_2N$ continues as aging progresses [54].

*Table 3. Derivatives of intra-granular $M_{23}(C, N)_6$ precipitates in SS 316LN*

| Sample | isolated cubes, | Primary stringers | Secondary stringers | clusters laths |
|---|---|---|---|---|
| N07 | √ | √ | × | × |
| N11 | √ | √ | × | × |
| N14 | √ | × (rare) | × | × |
| N22 | √ Primary stringers are more in comparison with isolated cubes | √ | √ | √ |

In present authors earlier study, while considering tensile properties, rise in hardness with nitrogen concentration is noticed from N07 to N14, however N22 steel doesn't show any significant variation [16]. Addition to the total increase in precipitate density, coarsening of incoherent precipitates ($Fe_2Mo/Cr_2N$) and cluster formation of coherent $M_{23}(C, N)_6$ stringers instigate depletion of strengthening elements in the matrix, and dislocation shearing through the precipitates, leading to the anomalous change in hardness of N22 steel. The mechanism of dislocation shearing through $M_{23}(C, N)_6$ precipitate in a mechanically tested sample has been imaged by diffraction contrast imaging in TEM under two-beam dynamical condition as shown in figure 17. Figure 17a depicts dislocation pileups in a slip band which smoothly penetrates through $M_{23}(C, N)_6$ of perfect cube morphology. The continuity of dislocation lines on the interface of matrix and precipitate indicates negligible strain field at the boundary which proves the extent of coherency between phases. Figure 17b is the diffraction pattern at two beam condition at which the bright field image has been recorded. Additionally, the size attained by the clusters of cubes is very large; up to hundreds of micrometers (Section 3.4), which might impart detrimental effect to the system.

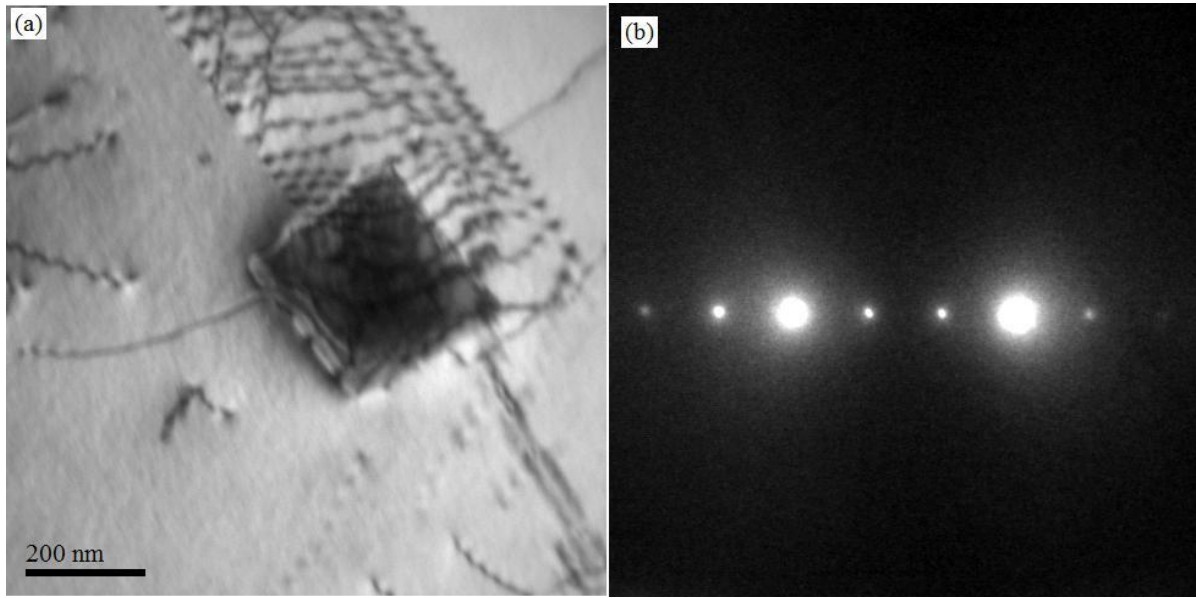

Figure 17. Dislocation shearing through $M_{23}(C, N)_6$ shows perfect coherency between the matrix and precipitates with negligible strain field at the interface: (a) Bright field TEM image under two-beam dynamical condition, and (b) Selected area diffraction pattern.

It is noted that the ultimate tensile strength increases with the increase in nitrogen content [16]. Excess addition of interstitial elements results in the formation of substantial secondary phases while aging gives a precipitate strengthened effect to the system. Hence the mechanical properties of the steels with varying nitrogen content can be interpreted more relatively with precipitate strengthening effects than the remaining strengthening mechanism. To summarize, the study demonstrates the role of nitrogen as a precipitate strengthener (indirectly) than an interstitial strengthener in SS 316LN during prolonged aging.

## 5. Conclusions

In long-term aged SS 316LN, the inviolate features, heterogeneous grain distribution by bimodal division and density of twins with varying nitrogen content (0.07, 0.11, 0.14 and 0.22 wt. %) has been visualized with light microscopy by applying discrete preparation methods in vibratory polisher.

Increase in precipitate density with nitrogen contentisa significant observation that emerges by bright field optical imaging. Additionally, the presence of distinct secondary phases has been identified by colour contrast and morphological features. Analysis of selected area diffraction patterns and amplitude contrast images in TEM reveals the identity of precipitates as $M_{23}(C, N)_6$ carbo-nitrides, $Fe_2Mo$ intermetallics and $Cr_2N$.

$M_{23}(C, N)_6$, and $Fe_2Mo$ form plentifully in all steels, whereas, $Cr_2N$ is largely distinguished in higher nitrogen-containing samples (N14 and N22). The contiguity of precipitates has been characterized further with diffraction contrast imaging in transmission electron microscopy. Moreover, the planar defects in $Fe_2Mo$ and hexagonal symmetry (hP11) for $Cr_2N$ have been established.

The distribution of morphological derivatives of $M_{23}(C, N)_6$ with nitrogen content has been categorized exclusively by optical imaging as four intra-granular varieties in addition to the grain boundary carbides. The documented intra-granular $M_{23}(C, N)_6$ are, (1) isolated cube-shaped precipitates, (2) primary stringers (3) secondary stringers or clusters of primary strings, and (4) clusters of preferentially grown laths along $\{111\}/\langle 110\rangle\gamma$. Each cube and branch of stringers are preferentially grown along $\{111\}/\langle 110\rangle$ with the cube on cube orientation relationship and coherent with the matrix γ. Thus the six crystallographic variants for the stringers and clusters are identified from the multiplicity of $\langle 110\rangle\gamma$. Prediction has been made for the three-dimensional formation of secondary stringers and lath clusters from three possible preferential $\langle 110\rangle$ directions in each $\{111\}$ plane.

The secondary stringers and clusters of laths are only visible in high nitrogen samples (N22), whereas, isolated cubes are evident in all samples. The stringers are hardly seen in N14 in comparison with the remaining samples might be due to the presence of a considerable population of contiguous precipitates of $Fe_2Mo$ and $Cr_2N$.

Structure-property correlation confirms the collective effect of precipitate density in tailoring the mechanical properties at a higher temperature such as yield strength, ultimate tensile strength, rupture strength, creep resistance, etc. The cluster formation in high nitrogen steel (N22) attribute to hardness variation by dislocation shearing mechanism.

**Acknowledgements**

Authors acknowledge Dr. A. K. Bhaduri, Director, IGCAR, Dr. Shaju K. Albert, Director, Metallurgy Materials Group, IGCAR, and Dr. S. Raju, Associate Director, MCG/MMG/IGCAR for constant encouragement and MRPU for providing the experimental facilities. Authors would like to express their sincere thanks to Dr. R. Divakar, Associate Director, MEG/MMG/IGCAR and Dr. Sruthi Mohan for technical discussions which are relevant for data interpretation.